\documentclass[12pt,reqno]{amsart}
\usepackage{graphicx,color}
\usepackage[centertags]{amsmath}
\usepackage{enumerate}

\begin{document}

\title[The Crisis Of Evidence]{The Crisis Of Evidence: Why Probability \& Statistics Cannot Discover Cause}

\author{William M. Briggs}
\email{matt@wmbriggs.com}
\urladdr{wmbriggs.com} 

\begin{abstract}
Probability models are only useful at explaining the uncertainty of what we do not know, and should never be used to say what we already know. Probability and statistical models are useless at discerning cause. Classical statistical procedures, in both their frequentist and Bayesian implementations are, falsely imply they can speak about cause. No hypothesis test, or Bayes factor, should ever be used again. Even assuming we know the cause or partial cause for some set of observations, reporting via relative risk exaggerates the certainty we have in the future, often by a lot.  This over-certainty is made much worse when parametric and not predictive methods are used. Unfortunately, predictive methods are rarely used; and even when they are, cause must still be an assumption, meaning (again) certainty in our scientific pronouncements is too high.
\end{abstract}

\maketitle

\section{Introduction}

What is the point of probability and statistics? Only {\it one} thing: to quantify or explain the uncertainty in that which we do not know. Probability is not needed to tell us what we already know. Surprisingly, many don't agree. 

Suppose we learned that 1,000 people were ``exposed" to PM2.5---which is to say, particulate matter 2.5 microns or smaller---at some zero or trace level, and that another group of the same size was exposed to high amounts---call these two groups ``low" and ``high PM2.5".  Suppose, too, it turns out 5 people in the low group developed cancer of the albondigas, and that 15 folks in the high group contracted the same dread disease.  What is the probability that {\it more} people in the high group had cancer?  

If you answered anything other than 1, you probably had advanced training in classical statistics. Given our observations, 1 is the right answer. What is the probability that 15 people in the high group had cancer? Same thing. What is the probability that 3 times as many folks in the high group had cancer? Again, the same. Easy.

We have demonstrated that we do not need probability models to tell us what we already know.

What are the real questions of interest here? Obviously: what {\it caused} the observed difference, and what we can we say about the uncertainty of {\it new} people having cancer given their exposure? Both of these are things which we do not know. Probability might be able to help us.  Yet here is where the trouble begins.  It is the conceit of classical statistics, in both its frequentist and Bayesian forms, to presume it can say what causes events. That probability is believed to have this power is the source of much and lasting grief.

This is a tale of woe.  So pervasive are the faults which I'll detail, that it's doubtful that you'll believe the extent of the problem. How could science have gone so wrong? Scientists are smart people! How could agencies such EPA, IPCC, FDA, {\it et cetera} come to rely on such faulty methods?  The answer is easy: classical methods confirm prejudices.

\section{Cause}

Some thing or things caused each unfortunate person in our experiment to develop cancer. What could this cause or these causes be? Notice I emphasize that there may be more than one cause present. It needn't be the same thing operating on each individual. Each of the 20 people may have had a different cause of their cancer; or each of the 20 may have had the same cause.  And this is so even though it may be that cancer of the albondigas is caused in the human body in only one way. Suppose some particular bit of DNA needs to ``break" for the cancer to develop, and that this DNA can only break because of the presence of some compound in just those individuals with a certain genetic structure. Then the cause or causes of the presence of this compound become our main question: how did it come to be in each of these people? That cause may be the same or different.

What is classical statistic's answer to cause? The hypothesis test. Let's review what that is. Hypothesis testing in frequentism or use of Bayes factors in Bayesianism are the sources of fallacy I mentioned. Hypothesis testing (or Bayes factors and decision analysis) is a procedure which should be abandoned with alacrity. 

All hypothesis tests follow the same course. Step 1 is to posit a (usually) parameterized probability model for the observations.  Step 2 is to form the so-called ``null" hypothesis. This is a statement that there is ``no difference" between the groups under consideration with respect to developing cancer of the albondigas. In our case, this would be the same as saying there is ``no difference" between the low and high PM2.5 groups. We'll come back to this strange ``no difference." 

Step 3 is to form some function of the observed data---this is called a {\it statistic} (hence the field of statistics).  Step 4 is where the magic happens. I mean this literally. Gigerenzer \cite{Giger2004} calls Step 4 ritual; I say it is mysticism, pure magical thinking. Step 4 calculates: given the null hypothesis is true, and given the model we chose, and given the data we saw, the probability of seeing a larger test statistic than the one we actually saw (in absolute value) assuming we could repeat the experiment an infinite number of times. This strange creature is called the p-value.

Everybody knows that if the p-value is less than the magic number, then success is yours. The observed difference is ``statistically significant"! Papers can now be written; grants applied for; beliefs have been validated.

Now I've told you what the p-value meant, and it'll be a sharp individual who could recount what I just said, the definition being so difficult; but what does everybody {\it assume} it means?  That higher PM2.5 {\it caused} the observed difference.

And that is false. 

There is no indication in the data that high levels of PM2.5 cause cancer of the albondigas. If high levels did cause cancer, then why didn't every of the 1,000 folks in the high group develop it? Think about that question. If high PM2.5 really is a cause---and recall we're supposing {\it every} individual in the high group had the same exposure---then it should have made each person sick. Unless it was {\it prevented} from doing so by some other thing or things.  And that is the {\it most} we can believe. High PM2.5 cannot be a complete cause: it may be necessary, but it cannot be sufficient. And it needn't be a cause at all. The data we have is perfectly consistent with some other thing or things, unmeasured by us, causing every case of cancer. And this is so even if all 1,000 individuals in the high group had cancer. 

This is true for {\it every} hypothesis test; that is, every set of data. The proposed mechanism is either {\it always} an efficient cause, though it sometimes may be blocked or missing some ``key" (other secondary causes or catalysts), or it is {\it never} a cause.  There is no in-between.  Always-or-never a cause is tautological, meaning there is {\it no} information added to the problem by saying the proposed mechanism {\it might} be a cause. From that we deduce a proposed cause, absent knowledge of essence (to be described in a moment), said or believed to be a cause based on some function of the data, is always a prejudice, conceit, or guess. Because our knowledge that the proposed cause only might be always (albeit possibly sometimes blocked) or never an efficient cause, and this is tautological, we cannot find a probability the proposed cause {\it is} a cause. 

Consider also that the cause of the cancer could not have been high PM2.5 in the low group, because, of course, the 5 people there who developed cancer were not exposed to high PM2.5 as a possible cause. Therefore, their cause or causes must have been different {\it if} high PM2.5 is a cause. But since we don't know if high PM2.5 is a cause, we cannot know whether whatever caused the cancers in the low group didn't also cause the cancers in the high group. Recall that there may have been as many as 20 different causes.  Once again we have concluded that nothing in the plain observations is of any help in deciding what is or isn't a cause. 

Given the multitude of possible measures we can make on people---everything from whatever they've eaten over the course of their life to the environments to which they have been exposed, and on and on almost endlessly---it is more than reasonable to suppose that we can discover some thing which is also different between the two groups besides exposure levels. Suppose it turns out---and something like this almost surely will---every person in the high group ate at least one more banana than did folks in the low group.  That means whatever conclusions we reach via our statistical analysis, whatever powers of discovering cause we ascribe to hypothesis testing,  we could have equally well put down to having eaten more bananas. This is because the label ``low PM2.5" and ``high PM2.5" can be swapped for ``low banana" and ``high banana", a set of measurements just as true and valid. If you are able to grasp this, you are able to see how deep the problem of classical statistics has dug.

It turns out that the p-value for this data, based on an ordinary test of ``difference" in proportions---a highly misleading name because we have already agreed that the proportions {\it are} different with probability 1---is 0.043. This is smaller than the magic number so we ``reject" the null hypothesis. That null was ``no difference" between groups with respect to cancer.  Rejecting this is to accept the belief, or to act as if, there {\it are} differences. But what would that possibly mean?

Does it mean that high PM2.5 {\it caused} the difference?  We have already seen this isn't so. Could it mean that high PM2.5 is merely ``associated" with or ``linked to" the difference?  But what could {\it associated} or {\it linked to} mean unless they were somehow causal, either saying PM2.5 is a direct cause or that it is part of the causal path, which is the same as saying PM2.5 is a necessary but not sufficient part of the cause, much the same as having (say) a certain genetic variant is a necessary but not sufficient cause of having some disease (for instance, sickle cell).

But this can't be either, because then eating extra bananas would also have to be ``associated with" or ``linked to" cancer in the same sense. So would every other thing we could measure that is different between the two groups. And this is absurd. 

Clearly, there was {\it some} thing or {\it some} things different between the two groups. There  must have been, because the number of people who got cancer was different, and the difference was caused, as must be true. But there is absolutely nothing in the observations themselves that tell us what this cause was or what these causes were. {\it Nothing.} This must be understood before we can continue. 

We are not just discussing PM2.5. The criticisms here apply to {\it every} classical statistical analysis ever done. The reader should also note that the {\it same} criticisms apply if instead of hypothesis tests we looked at Bayesian posterior distributions or Bayes factors. The same dismal over-confidence is present. The only advantage Bayes brings is coherence in discussing probability. It is no way fixes the misunderstanding of causality. 

\section{Negative Positivism}

Suppose one fewer person in the high group had cancer: only 14 got it, and still 5 in the low group. The p-value now is 0.065.  What is the classical decision?  To ``fail to reject" the null hypothesis. Notice that we never ``accept" the null; we just ``fail to reject" it.  Have you ever wondered about that bizarre terminology?  Well, it is the product of the same mistake that led to misunderstanding the nature of causality. 

Let's play a game of {\it Who Said It?}:
	\begin{enumerate}[{(}a{)}]
		\item ``We have no reason to believe any proposition about the unobserved {\it even after} experience!"
		
		\item ``There {\it are} no  such things as good positive reasons to believe any scientific theory."
		
		\item ``The truth of any scientific theory is exactly as improbable, both {\it a priori} and in relation to any possible evidence, as the truth of a self-contradictory proposition" (i.e. It is impossible.) 
		
		\item ``Belief, of course, is never rational: it is rational to {\it suspend} belief."
	\end{enumerate}
The first is from the grandfather of skepticism, David Hume \cite{Hum2003}.  The others are all Karl Popper \cite{Pop1959,Pop1963}.   It was Hume who took skeptical views about causation to their limit. To Hume, we can {\it never} know, never really be certain, of {\it any} cause. All events we see are just this-follows-that, our awareness was of recurring patterns which he called ``constant conjunctions", happenstance occurrences. What we took to be cause was just our prejudice overlaid on events occurring close in time, which to Hume were really ``entirely loose and separate." 

Hume's views were and are highly influential, more among philosophers than scientists directly. Hume particularly influenced the logical positivists and they {\it did} have a lot of pull with scientists. Karl Popper was one of the leading positivists.  Popper really did think that you could never believe any scientific theory, in the sense of knowing it with certainty. But he did say that you could disbelieve one, which is to say, know with certainty a theory was in error. This led to his theory of falsifiability. Theories were not ``scientific" unless they were falsifiable, which is to say, could be refuted by empirical observation. But then all math, including nearly all probability, is thus non-scientific because no math and most probability statements can be falsified. Now the logical positivists were infamous for claiming only empirical knowledge counted, a statement which itself cannot be verified empirically. This led philosopher David Stove \cite{Sto1983} to call the logical positivistic interregnum in the Twentieth Century philosophy an episode in black comedy.

The point for us is that RA Fisher (geneticist and statistician) was keen on Popper's idea of falsifiability and wanted to build it in to statistical practice \cite{Fis1973b}. This he did with p-values. This is why we ``reject" a hypothesis---what Fisher thought was ``practically" falsifying it---but why we never accept one. Accepting the null, i.e. believing or acting as if it were true, would be believing we knew a scientific hypothesis, which would be irrational. Thus we can only ``fail to reject" the null. 

Popper's views are, as you might have guessed, self-contradictory, as was the foundational dictum of logical positivism. If you reject a proposition and say it is false, then you are accepting its contradictory. Which is to say, you are believing in the truth of a proposition, which Popper said is always irrational.   

Add to that contradiction that we are never falsifying anything with a wee p-value, and you have a dicey philosophy upon which to build scientific inference. You are not falsifying with a p-value because a p-value does not say any proposition is true or false. Indeed, a p-value says almost nothing at all. It says something about the probability of seeing values of test statistics given some assumptions, and that's it. It certainly says {\it nothing} about cause. P-values are an act of will; they are not scientific or objective.

What happens when we accept---nobody really thinks in terms of failing to reject---a null hypothesis? We say the results are not ``statistically significant", of course, but we also say the results are ``due to" chance or randomness. {\it Due to} means caused. But chance is not a cause; neither is randomness. Chance never causes {\it anything}. Chance and randomness aren't material, they aren't forces, and thus cannot be causes. 

And then it could be that every case of cancer in the high group was certainly caused by high PM2.5, while every case in the low group was caused by something else. Thus to say rejecting the null {\it proves} that PM2.5---or anything else---cannot be the cause of cancer is to commit a fallacy. Hypothesis testing is nothing but wishful thinking.

\section{The Essence and Power of Cause}

What is the solution? Purge the notion that probability and statistics can prove agency and come to a new view of causation. And by this I mean return to the old view, the Aristotelian view, the view most physical scientists actually hold, or used to, when they come to think of causality. Although this is a huge and difficult subject, we can boil it down to essence. We need to understand the essence, the whatness or quiddity, of the objects which are thought to be causally related. 

How do we know gravity {\it causes} the apple to fall? Not because some instrumentalist equation tell us so, but because we understand it is the nature of gravity to bend space, or simply because we understand it is the power of gravity to pull or the nature of things to fall. Math, while extraordinarily helpful, is not physics; it is not biology. It can help bound or quantify effects, but it can't determine what essence is. Just as with probability, excessive reliance on math leads to the Deadly Sin of Reification, the belief that equations come alive and are objectively real.

Essence is of the essence. Just one out of an inexhaustible supply of examples. As of this writing, there are 1,446 living Grand-masters of chess. Only 33 of these aren't men. Grand-master Nigel Short suggested this ``disparity" was be{\it cause} it was the nature of men to enjoy the game more and be{\it cause} men possessed more talent.  Amanda Ross, a chess writer, disagreed and said sexism was the cause.\footnote{http://wmbriggs.com/post/15778/} Now whichever side of the debate you prefer, understand that the data, the plain data, equally well supports both sides. There would nothing gained, but much lost, in any attempt to model this situation. Besides these two possibilities, it is possible to think of many other possible causes. Cause is {\it under-determined} by the data. The only way to solve this, or any problem, is to grasp the nature or essence and powers of the things involved.

There are a growing number of philosophers calling for a return to this old way, to the understanding that cause has four dimensions---formal, material, efficient, and final---to return to metaphysical realism. Philosophers like David Oderberg \cite{Ode2008}, Edward Feser, Nancy Cartwright, and others, all of whom should be consulted. (This brief introduction cannot do the subject justice.) 

In the case of things like whether high PM2.5 causes cancer or some other disease, we need to do the obvious. Posit {\it how} and then {\it verify directly}. It's a simple as that---philosophically speaking. It is extraordinarily difficult to do in practice, naturally. We need to understand the etiology of the disease under consideration; that is, the nature of cancer or other disease and the exact pathways the disease takes, the nature of the human body in its response to dust, the differences in bodies and behaviors that explain why some are susceptible and some not, the power of dust to cause disease and the body to fight it off, the environment in which dust is transported, and so on. A monumental task when the causes are complex, as they surely are for cancer and heart diseases. 

And this is done in all the ways we already know about. By brutal hard work. Bench and lab work, painstaking testing, animal experiments, genetic analysis, and on and on. By making predictions {\it and then waiting to verify them}. So difficult, time-consuming, and expensive is this kind of work that relief is sought. Easy answers are swapped for hard ones. Statistics provides these easy answers by pretending to be much more than it is.  The fallacy of ``We Have To Do Something" is ever harkened to. Action is desired, and justification for the action is needed. Classical statistics provides it.  It all sounds so scientific!

\section{The Real Dismal Science}

We have seen that the only reason to use probability and statistics is for one thing: to quantify or explain the uncertainty in that which we do not know (not all probability is quantifiable).  One thing we often do not know is what causes an effect. We have seen that probability and statistics can never prove what causes what, though it might be able to put a measure on the likelihood that one cause over another was operative. But probability can only be comparative. We have to know what the possible causes are {\it in advance} before probability is of any use. If we don't have any idea what a cause was, probability is useless. 

And when we {\it do} know a cause, we do not need probability. The classic example is the gambler's fallacy. We observe red has come up on the wheel the last 10 times. Black is not therefore `due", because why? Because we know or assume the {\it causes} of the roulette wheel have not varied. This is an excellent example because it shows any statistical model based on observed data applied to roulette wheel data is bound to be wrong.  If you don't like mechanical roulette wheels, think of quantum mechanical events, composed of parts that do not wear and that always act the same, and which are highly predictable. And they are so even though we do not (in most cases) know the ultimate cause of QM events.  

Why is probability useful for QM events when we do not know the cause? Probability, encapsulated in such formulas like Schr\"odinger's equation, predict very well, but they do not tell us the cause of why, for instance, the electron in a two-slit experiment takes the top and not the bottom slit. Yet we can use probability models to make excellent predictions. {\it Not} as excellent, you must understand, as causal models which would predict without error. But, still, QM probability calculations are good because of two reasons. One, they have found the right conditioning premises, the circumstances which must be observed for the experiment, and, two, they have been {\it verified to work.}  

That last point is the most crucial. The (quantum mechanical) standard model of physics would have been abandoned long ago had it not been demonstrated to work, in the sense that it has made useful, verifiable and verified predictions. This is not how statistics works, however, and it's increasingly not how science works. It used to be that if a theory made lousy predictions, the theory was tossed out on its ear. But now a theory's desirability counts more in its favor than anything else. Its ability to match reality is welcome but in no way required. I need only mention global warming to prove this point.  That we have reached this dismal situation is in good part due to statistics and the mistaken belief that it can show cause.

Here's what should happen in the absence of knowledge of cause. Like in QM, the conditions which are thought most related to the cause of the event of interest should be measured and noted. The events themselves should also be measured. To the largest extent possible, given our current knowledge, the relations between these measures should be deduced.  Pure deduction is usually only possible on the crudest or largest scales. Whatever relationships are left over can be modeled probabilistically. But the model should not be relied upon in its explanatory sense, as it is in both classical frequentist and Bayesian statistics. It should be used in its predictive sense, like it is in physics and engineering and, to a smaller extent, in medicine. Once the predictions are verified, then we know we have a useful model. Until then, {\it we do not know the model is useful and it should not be used to make decisions.}  Lastly, as emphasized in the roulette wheel, we should {\it not} use a probability model when we know the cause or understand the essence. 

This represents a fundamental, really a gargantuan, shift in the practice of science. It will be costlier and slower than the current way, but it will also be vastly superior. And it will eliminate the pandemic of over-certainty we now experience. 

Let's see how this might look using the PM2.5 example.

\section{Fallacy}

There is plausible suspicion that PM2.5 {\it might} cause disease (for brevity's sake, I'll only speak of cancer). We know this because we suspect it is in the nature of fine particulate matter to interact with, and possibly interfere with, the functioning of the lungs, the nature of which we also have some grasp. We do not {\it know}---and never forgot that we can only know what it is true: though we can {\it believe} anything---that PM2.5 causes cancer. A reasonable condition, given what we have learned from other dose-response relationships, is that greater exposure to PM2.5 will give more opportunity for whatever it is in PM2.5 that causes cancer to operate.

The proper conditions upon which to build any model would be the actual exposure of PM2.5, as well as those other things we know about the human body and behavior we suspect are related to the disease in question.  We take those measurements and also tally mortality, morbidity, disease state, or whatever else is of interest and then form a statistical model.  There will be very little to no deduction in this approach, however, which means we must rely even heavier on our model's predictive ability.

Notice that in this approach we {\it must} assume that (high) PM2.5 is {\it always} a cause but that sometimes it is stopped from operating because of some lack: say, a person has to have a specific genetic code, or must inhale the dust only when breathing is labored, or some chemical must be present, or whatever (the exact conditions may be exceedingly complex). As we saw above, the only other assumption is that PM2.5 is {\it not} a cause, and if it is not, then we must not use a probability model. That would be to commit the gambler's fallacy, as in the roulette wheel.

Is this what is done? No. 

Instead, the usual practice is to rely on proxies of exposure---and {\it not} just for PM2.5. A typical abuse is to measure an average value or posit a ``land-use model" of PM2.5 and then suppose everybody in or near some locale was exposed to a predicted level of particulates.  This is silly, of course, and is what I call the epidemiologist fallacy. It's related to the ``ecological fallacy", but I prefer the former name because without this fallacy, most epidemiologists, especially those employed by the US government, would be out of a job.  The epidemiologist fallacy is also richer in content than the ecological; it occurs whenever an epidemiologist says, ``X causes Y" but where he never measures X {\it and} where he uses classical statistics to claim proof of a cause. 

You might think such egregious practices are rare. You would be mistaken. So common is this error, so entrenched is this pseudo-scientific practice, that it forms the bulk of research conducted on things like PM2.5 or other environmental exposure.  Indeed, I have yet to see a paper which claims PM2.5 causes disease that didn't use it.  For examples see e.g. \cite{Jer2011,RaaAnd2013,DocPop1993,PopThu1995,StiChe2015,PopTur2015,HooWan2013}. Also see \cite{Bri2014} for details on the epidemiologist fallacy.

Still, the error is really only a major problem for the classical (frequentist or Bayesian) statistician who uses the hypothesis test or Bayes factor. He will be incredibly over-certain in is pronouncements.  But the predictive statistician does not suffer these faults. 

\section{PM2.5 Specifically}

This current work is concerned mainly in the nature of evidence, but for those interested in PM2.5 specific, an indispensable paper, little cited, is by John Gamble ``PM2.5 and Mortality in Long-term Prospective Cohort Studies: Cause-Effect or Statistical Associations?" \cite{Gam1998}. I'll only cover a few highlights, but the paper is a must read for all interested in this subject. Gamble says everybody looks to the (cited above) Six Cities Cohort and the American Cancer Society Cohort\footnote{I applied for access to these data to apply predictive techniques but was rejected.} which discovered statistical significance of mortality or morbidity, but the Seventh Day Adventist cohort, which did not find significance, is routinely ignored.  Gamble also understood the biases involved in using ecological and not actual measurement data. He correctly notes ``exposure must precede disease" but that in the two main studies ``estimates of exposure meet neither of the temporal criteria for latency or precedence."

It was found in the Six Cities Cohort that average values of PM2.5, often used as a proxy to exposure, have very little to no correlation to actual exposure levels, that ``Data from other cities also show that outdoor concentrations are poor surrogates for personal exposure", and that ``ambient PM is not correlated with nonambient PM".   Lifetime exposure is also never or badly measured, especially considering modern mobility.

Perhaps the most (unintentionally humorous) finding was that the two main cohorts estimated mortality risk of about 2 to 2.3 for 25 pack-year smokers. Now smokers receive orders of magnitude greater exposure, both in dose and length, to PM2.5 than do non-smokers.  Yet non-smokers with an estimated exposure of 20 $\mu g/m^3$ over ambient PM2.5 (about 1-20 $\mu g/m^3$) had mortality risk ratios from 1.17 to 1.26. Smokers receive about 20,000 $\mu g/m^3$ {\it annually}. Assuming increasing dosage are worse, Gamble worked out that this necessarily implies that PM2.5 is about 150 to 300 times {\it more toxic} than cigarette smoke! We can blame the curse of statistical significance for this one.

What about understanding the nature of the disease. Gamble was writing in 1998 and said there ``appears to be general agreement that no plausible mechanism is presently available to explain the associations between chronic exposure to PM2.5 air pollution and increased mortality." Further, ``experimental data of lifetime exposure of animals to fine particulate matter showing no increased mortality even though exposures are so high as to produce lung overload".

Incidentally, the EPA on their information page,  does not mention non-man-made sources of PM2.5.\footnote{http://www.epa.gov/pmdesignations/faq.htm}

\section{Futility Of Risk}

A common way to report on these matters is by relative risk. The simplest way to think of it is this: the probability of having the disease, or of dying or of having whatever outcome of interest, given the high exposure, divided by the probability of disease given low exposure.  A relative risk of 2 means the disease is twice as likely to be found in the high group.  Relative risk is a misleading way to report, however, because it exaggerates risk, as I'll prove.

What's the difference between a probability of one in ten million and one of two in ten million?  The official answer is ``Not much," but the relative risk is 2, which is considered high.  If the two numbers were the (conditional on evidence) probability of disease in the low and high groups, one could honestly write a paper that said, ``Exposure to high PM2.5 doubles the risk of cancer."  But it would highly misleading.

What counts as ``exposed"?  Daily inhalation? If so, for how long a period and in what quantities? And what about the people ``not exposed (to high PM2.5)"?  They were ``not exposed" because they were {\it certainly different} than the people who {\it were} ``exposed." That they must be different is a logical truism: if they were the {\it same} they would have been exposed to high PM2.5, too. Since the low group were not exposed, they are different. In how many ways are the people in the two groups different?  Nobody knows, and nobody {\it can} know. The number of measurable differences, as we already saw, is astronomic.

Is our job made easy by defining ``exposed to" as ``any exposure of any kind in any high amount", which is certainly plain enough?  No, because it begs the question why the folks who did not see any exposure of any kind in any amount didn't. Were they off on holiday? Did they eat different foods? Come from a different culture? Were they older or younger? Have variant genetics?  You can go on and spend a lifetime and never be sure of identifying all things different between the groups.

Those problems are too hard, so let's do something as easy as possible, which will still lead to an important twist. Suppose God Himself told us that the probability of cancer in the exposed (in any way to high PM2.5) group is 2 in 10 million and the probability of cancer in the not-exposed group is 1 in 10 million. The relative risk is 2. 

Evidence suggests that Los Angeles is a ``hot spot" for high PM2.5 exposure. Now L.A. proper has about 4 million residents, some of whom will have been exposed, some not. Suppose for the sake of argument the population is split: half are exposed, half not. Remember: the Lord Himself assured us the relative risk of 2 is correct.  The chance is thus 67\% that {\it nobody} in the exposed group will develop cancer. Pause here. That's {\it no}-body. Meaning there's a two-out-of-three probability that not a single soul of the 2 million exposed people will get cancer. Anti-intuitive? Shouldn't be: even with a relative risk a whopping 2.0, there's still only a paltry 2 in 10 million chance of disease for any individual. 

Likewise, we can calculate the chance nobody gets cancer in the not-exposed group: it's 82\%. Isn't that  67\% versus 82\% odd? In the exposed group, the chance at least one person gets cancer is one minus the chance nobody does, or 33\%; and the chance at least one person gets it in the non-exposed group is thus 18\%. The risk ratio is now 1.8, a big change from 2! But how can this be when God Himself told us the relative risk is 2?  That's because the ``2" is an abstract number. When we start applying it groups of real people, the stark difference begins to disappear. For example, if we applied the same calculations to the population of New York (about 8 million) the real relative risk is 1.67. Lower still.  By the time we consider the population of the entire United States, the real relative risk dwindles to 1.

Why is this happening? Because relative risk is stated in terms of {\it single people}. When you're concerned only with yourself, it's the right calculation---but a silly one, because you're better off knowing the probabilities involved. Consider that you have the same relative risk moving from 2 in 10 million to 1 in 10 million as from 1 in 2 to 2 in 2, yet the situations are worlds apart!   

Let's think again about the population of LA. The chance that {\it just one} person of the 2 million in the exposed group develops cancer is 27\%.  That means there's a 94\% that either nobody gets it or just one person in 2 million does. And there's a 16\% chance just one person in the non-exposed group gets it, or a 98\% chance 0 or 1 people get cancer. The overwhelming probability---94\%---is {\it no more than one} poor soul develops cancer in the exposed group, and that it's more likely {\it nobody} does (67\%). The same is true in the not-exposed camp.  

Here's another way to look at it. In all of LA's population, some people in each group might get cancer. The chance that {\it more} people develop cancer in the exposed group than in the non-exposed group is only 28\%. Which is not even close to 100\%.  Even with a relative risk of 2, there's only just over a 1 in 4 chance of larger numbers of cancer patients in the exposed group.  Reflect on this. 

The largest chance, 59\%, is that both groups will have the {\it same} number of people who develop cancer. And there is even a 13\% chance that the not-exposed group will have {\it more people} who develop cancer than the exposed group!  There is an enormous difference between sensational relative and sober absolute risk. 


Does 1 in 10 million seem low ?  It doesn't to the EPA. In one of many guides (e.g. p. 5\footnote{http://www.epa.gov/reg3hwmd/risk/human/rb-concentration\_table/chemicals/SSG\_nonrad\_technical.pdf}), they fret over risks as tiny as that. However, they often reference risks of 1 in a million and 1 in 10 thousand as regulation-worthy. Let's use the latter and see what boosting the exposed cancer chance to whopping 2 in 10 thousand does to our calculations. 


We first need a workable relative risk---2 is really too large. We need one considered by the regulatory establishment high enough to warrant hand-wringing.  Use 1.06, the high-water relative risk in a series of widely touted papers by Michael Jerrett and co-authors \cite{Jer2011} who measured exposure to dust. Jerrett spoke of many diseases, mainly cardiac and respiratory ailments, but what matters here is the size of relative risk touted as worrisome. We're still claiming that all that counts is exposure or non-exposure; all or nothing. We can't do here the harder problem of multiple levels of exposure and the reasons for differences in exposure.  Assume some people had ``enough" exposure to dust and others didn't. The disease for the sake of illustration is still cancer, but it could be anything.


With a relative risk of 1.06, the chance of disease in the not-exposed group is 0.000189.  Here is a picture of the probabilities for {\it new} disease cases in the two groups in LA, still considering a population of 4 million and split exposure.

\begin{figure}[tb]
    \begin{center}
        \scalebox{.4}{\includegraphics{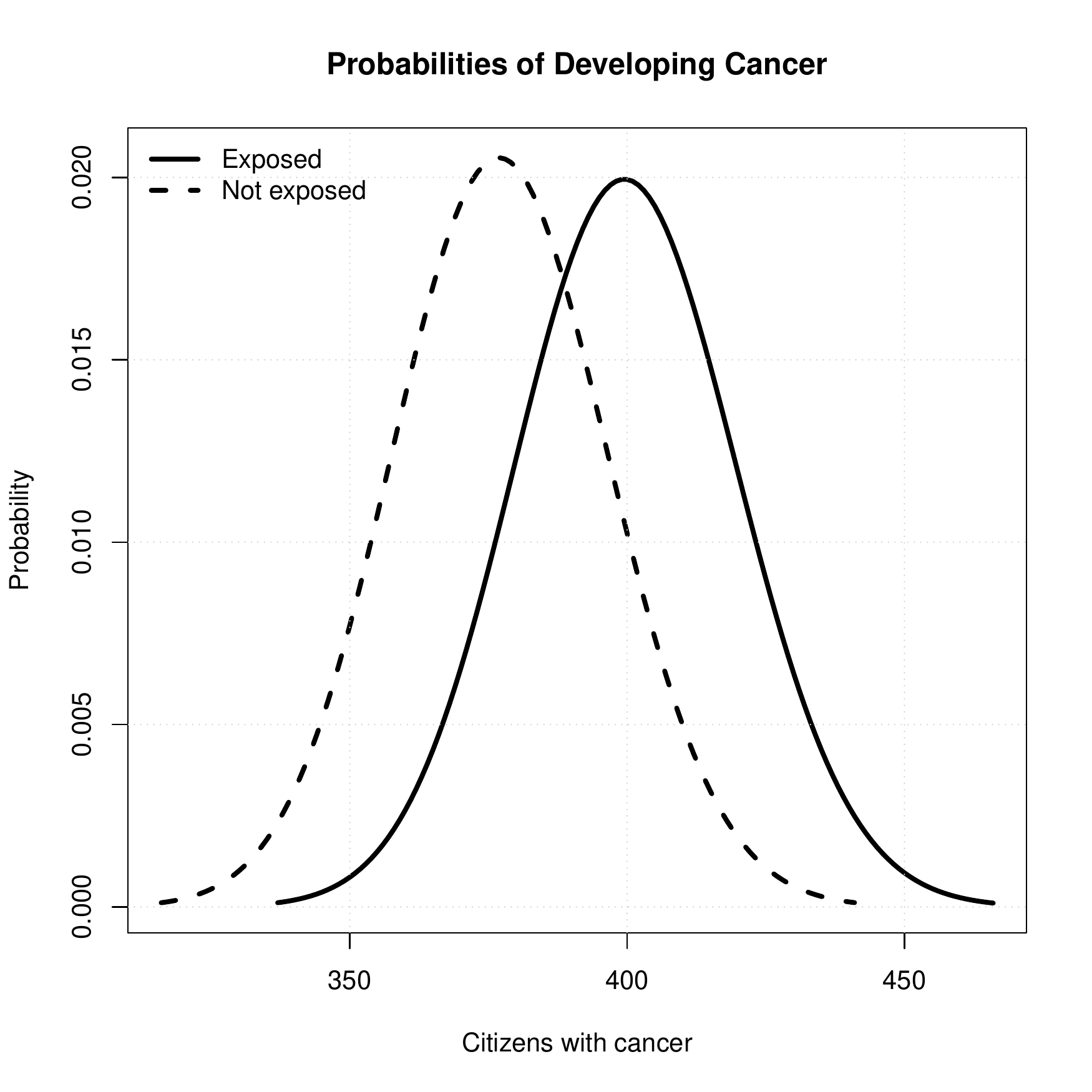}}
    \end{center}
\caption{\label{fig1} The probability citizens of Los Angeles develop cancer with and without exposure to high PM2.5. These are not normal distributions, but binomial and only appear smooth. These are non-zero probabilities for observable events.}
\end{figure}

There's a 99.99\% chance that from about 300 to 440 not-exposed people will develop cancer, with the most likely number (the peak of the dashed line) about 380.  And there's a 99.99\% chance that from about 340 to 460 exposed people will develop cancer, with the most likely number about 400.  A difference of about 20 folks.  Surprisingly, there's only a 78\% chance that {\it more} people in the exposed group than in the not-exposed group will develop cancer.  That makes a 21\% chance the {\it not-exposed} group will have as many {\it or more} diseased bodies. 

This not-trick question helps: how many billions would you pay to reduce the exposure of high PM2.5 to zero?  How many people would get cancer?  The answer is: {\it this statistical model says nothing about what causes cancer}.  As shown above, we first have to {\it assume} PM2.5 is a cause. We {\it can} say is that eliminating high PM2.5 eliminates high PM2.5, which is equivalent to saying that  everybody else, all 4 million folks, would be exposed to low PM2.5. Calculations show there's a 99.99\% chance that anywhere from about 650 to 850 people would get cancer, with the most likely number being around 760. 

\begin{figure}[tb]
    \begin{center}
        \scalebox{.4}{\includegraphics{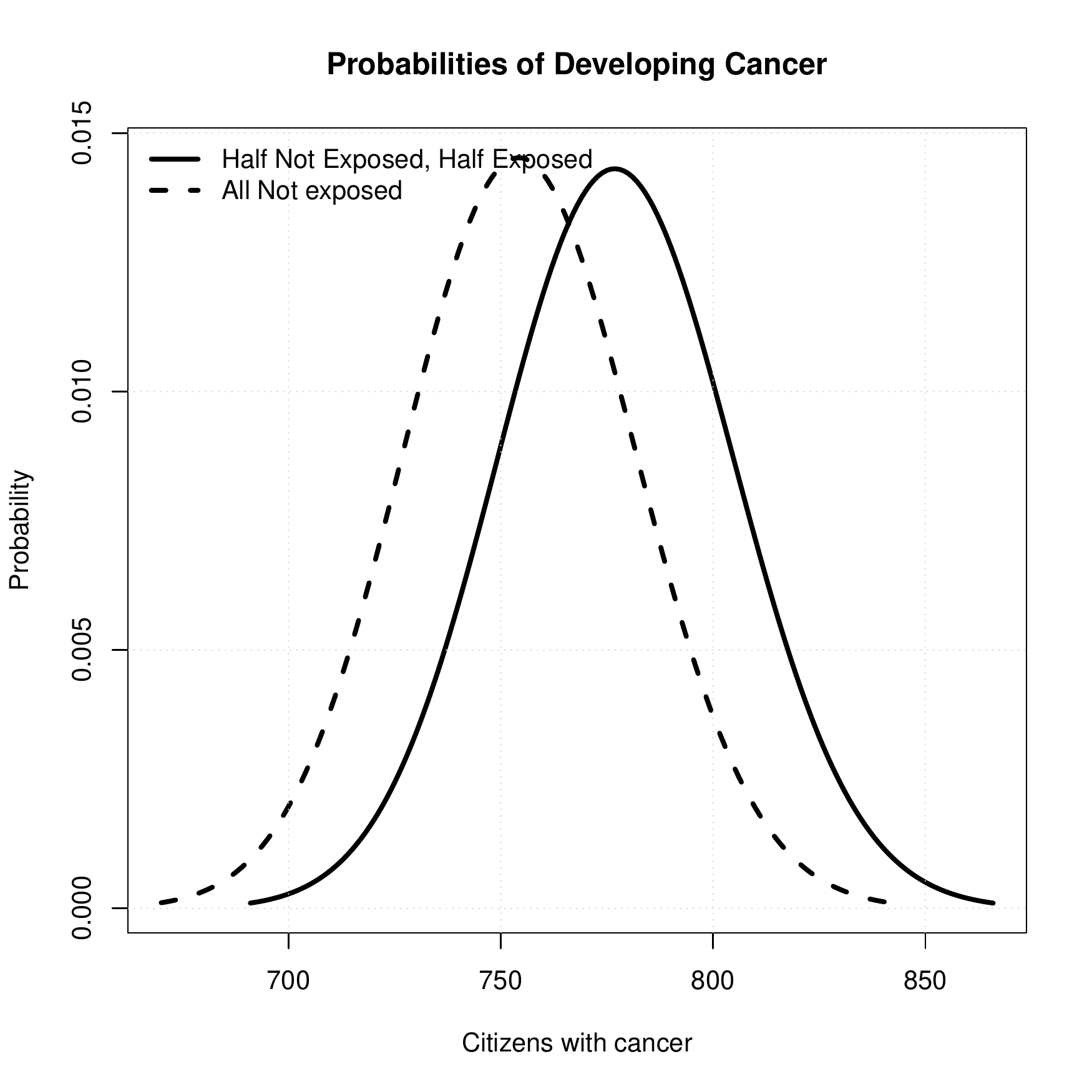}}
    \end{center}
\caption{\label{fig3} The probability of the number of people having cancer, when half the population of LA is exposed and half not, compared to the supposing the entire population isn't exposed.}
\end{figure}

There's a 99.99\% chance that from between just under 700 to just over 860 people will have cancer, given exposure is split in the population between high and low PM2.5. And there's the same chance that from about 670 to 840 people develop cancer assuming nobody is exposed to high PM2.5.  The most likely number of victims for the split population is 777, and it's 754 in the all-low population, again, only about 20 folks different.  There's only a 71\% chance more people in the split population would have more people with cancer than in the all-low scenario, but there's also a 28\% chance more people in the all-low scenario have cancer.  

This and not the previous picture is the right way to compare real risk---when the risk is known with certainty.  It's not nearly as frightening as the normal methods of reporting.

This figure shows there's a cap, a number which limits the amount of good you can do (regardless whether cancer is caused by PM2.5 or something associated with it).  If we use the 99.99\% threshold (adding a few 9s does not change the fundamental conclusion) and we eliminate {\it any possibility} of high exposure, then the {\it best} we could save is about 200 lives---assuming, too, the cancer is fatal. That comes from assuming PM2.5 is cause, that 866 exposed people develop cancer and 670 not-exposed people get it (the right-most and left-most extremes of both curves). There's only a 0.005\% chance that 866 or more exposed people get cancer, and there's a 99.985\% chance at least as many as 670 in the not-exposed people get it.  The {\it most likely} scenario is thus a saving of about 20 lives. Out of 4 million. Meaning our Herculean efforts to eliminate all traces of dust at {\it best} we'd affect about 0.004\% of the population, and probably more like 0.0005\%. And never forget we only assumed PM2.5 was a cause. If it isn't, all our efforts are futile. How many billions did you say?

There are other strong assumptions here. The biggest is that there is {\it no} uncertainty in the probabilities of cancer in the two groups. {\it No} as in {\it zero}.  Add {\it any} uncertainty, even a wee bit, and that expected savings in lives goes down.  In actual practice there is plenty of uncertainty in the probabilities. The second assumption is that everybody who gets cancer dies. That won't be so; at least, not for most diseases. So we have to temper that ``savings" some more. 

Assumption number three: exposure is perfectly measured and there is no other contributing factor in the cancer-causing chain different between the two groups. We might ``control" for some differences, but recall we'll {\it never know} whether we measured and controlled for the right things. It could always be that we missed something.  But even assuming we didn't, exposure is usually measured with error, as we have seen.  In our example, we said this measurement error was zero.  In real life, it is not; and don't forget Jerrett relied on the epidemiologist fallacy. Add any error, or account for the fallacy, and the certainty of saving lives necessarily does down more.

Let's add in a layer of typical uncertainty and see what happens. The size of relative risks (1.06) touted by authors like Jerrett get the juices flowing of bureaucrats and activists who see any number north of 1 reason for intervention.  Yet in their zeal for purity they ignore evidence which admits things aren't as bad as they appear. Here's proof.

Relative risk estimates are of course produced by statistical models, usually frequentist. That means p-values less than the magic number signal ``significance".  Now (usually arbitrarily chosen and not deduced) statistical models of relative risk have a parameter or parameters associated with that measure.  Classical procedure ``estimates" the values of these parameters; and as we've seen, the guesses are heavily model and data dependent.  Change the model, make new observations, and the guesses change. 

There are two main sources of uncertainty (there are many subsidiary). This is key. The first is the guess itself. We thus far assumed there was {\it no uncertainty} of the first kind. We {\it knew} the values of the parameters, of the probabilities and risk. God told us! Thus the picture drawn was the effect of uncertainty of the second kind, though at the time we didn't know it. We saw that even though there was zero uncertainty of the first kind, there was still tremendous uncertainty in the future.  Even with ``actionable" or ``unacceptable" risk, the future was at best fuzzy.  Absolute knowledge of risk did not give absolute knowledge of cancer. 

This next picture shows how introducing uncertainty of the first kind---present in every real statistical model---{\it increases} uncertainty of the second.

\begin{figure}[tb]
    \begin{center}
        \scalebox{.4}{\includegraphics{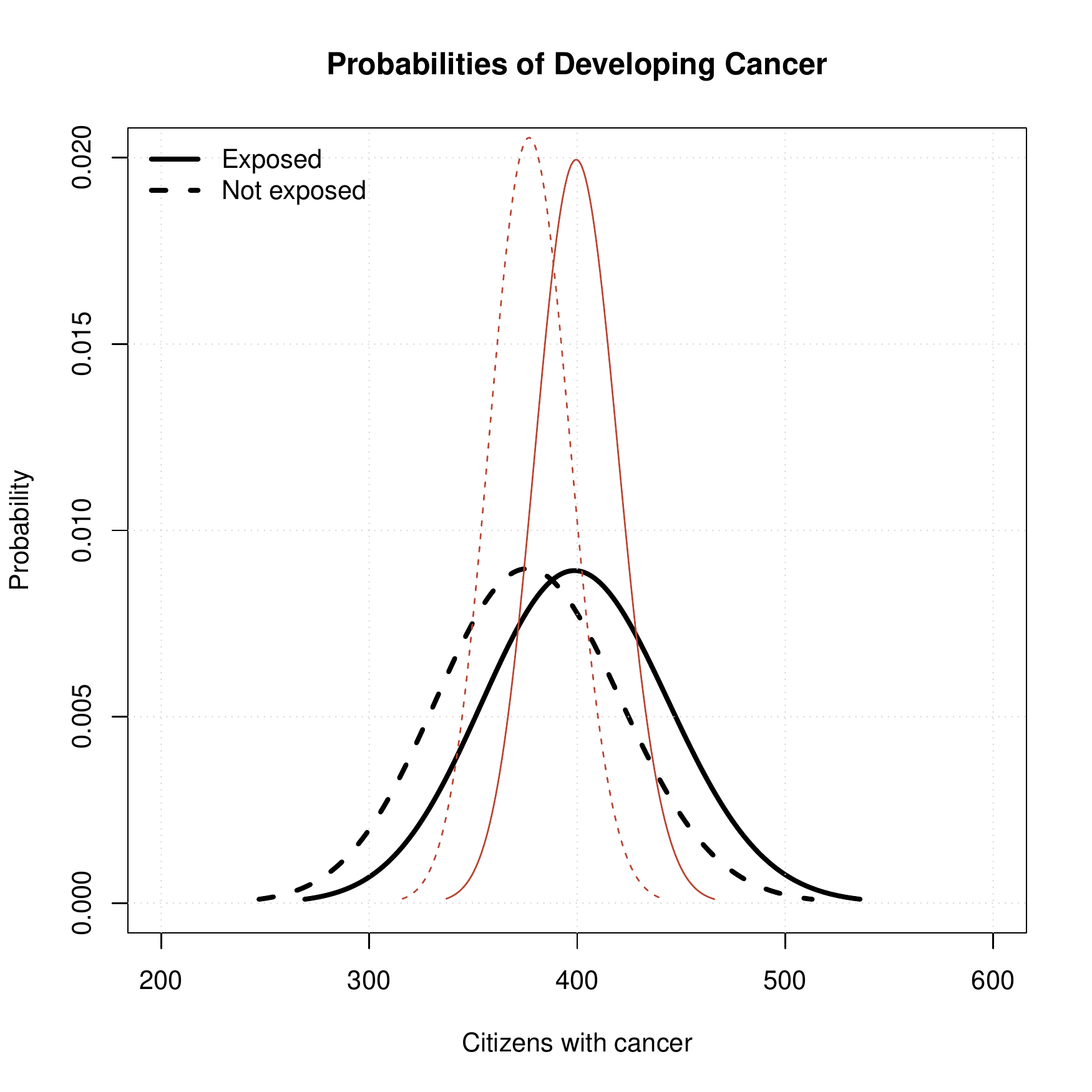}}
    \end{center}
\caption{\label{fig2} The probability citizens of Los Angeles develop cancer with and without exposure to high PM2.5, factoring in parameter uncertainty.}
\end{figure}

The narrow reddish lines are repeated from before in Fig. \ref{fig1}, which were the probabilities of new cancer cases between exposed and not-exposed LA residents assuming perfect knowledge of the risk. The wider lines are the same, except I've added in parameter uncertainty (since I don't have Jerrett's, this is only a reasonable guess).  In Bayesian terms, we ``integrate out" the uncertainty in parameters and produce the posterior predictive distributions. The spread in the distribution doubles: {\it uncertainty increases dramatically}.

There is also more overlap between the two black curves.  Before, we were 78\% sure there would be more cancer cases in the exposed group.  Now there is only a 64\% chance: a substantial reduction.  Pause and reflect.  Parameter uncertainty increases the chance to 36\% (from 22\%) that any program to eliminate {\it PM2.5} does {\it nothing}, still assuming PM2.5 is a cause. Either way, the number of affected citizens remains low.  Affected by cancer, that is. {\it Everybody} would be effected by whatever regulations are enacted in the ``fight" against {\it PM2.5}. And don't forget: any real program cannot eliminate exposure; the practical effect on disease must always be less than ideal.  But the calculations focus on the ideal.

As above, here are the real curves we should examine:

\begin{figure}[tb]
    \begin{center}
        \scalebox{.4}{\includegraphics{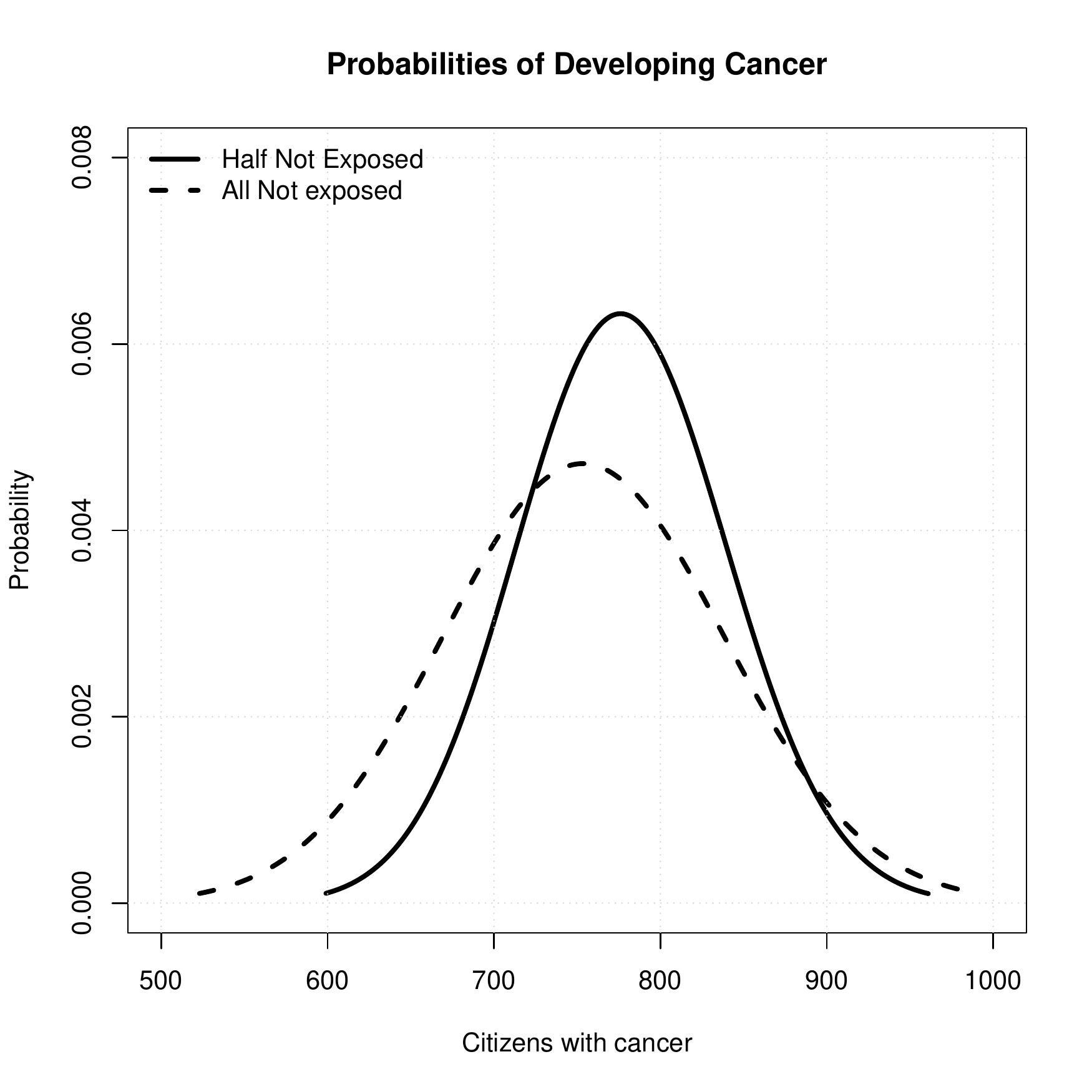}}
    \end{center}
\caption{\label{fig4} The probability of the number of people having cancer, when half the population of LA is exposed and half not, compared to the supposing the entire population isn't exposed, factoring in parameter uncertainty.}
\end{figure}

Isn't that amazing! By properly accounting for uncertainty, there is now only a 58\% chance that more citizens would develop cancer in the spilt-exposure group than in the all-not-exposed group. And there is a 41\% chance that more people would have cancer in the all-not-exposed group. There is more uncertainty in the all-not-exposed group, too, meaning the number of lives saves is up in the air; almost unknowable.  This is the benefit of examining predictive uncertainty. It gives the best, of course still model and data dependent, picture of what we can expect.

We're not done. We still have to add the uncertainty in measuring exposure, which typically is not minor. For example, Jerrett (2013) assumes air pollution measurements from 2002 effected the health of people in the years 1982-2000.  Is time travel possible? Even then, his ``exposure" is a guess from a land-use model, i.e. a proxy and not a measurement. Meaning he used the epidemiologist fallacy to supply exposure measurements. 

I stress I did not use Jerrett's model---because I don't have it. He didn't publish it. The example here is only an educated guess of what the results would be under typical kinds of parameter uncertainty and given risks and exposures. The direction of uncertainty is certainly correct, however, no matter what his model was. 

And there are still sources of uncertainty we didn't incorporate. How good is the model? (The only ``true" model is a fully causal one.) Classical procedure assumes perfection. But other models are possible. What about ``controls"? Age, sex, etc. Could be important. But controls can fool just as easily as help, as we now know. Anyway, controls don't change the interpretation. If there were controls, we would just have different pictures for each subset (say, males versus females, etc.).

All along we have assumed we could eliminate exposure completely.  We cannot. Thus the effect of regulation is always less than touted. How much less depends on the situation and our ability to predict future behavior and costs. Not so easy.

I could go on and on, adding in other, albeit smaller, layers of uncertainty. All of which push that effectiveness probability closer and closer to 50\%.  But enough is enough. You get the idea.

\section{Conclusion}

Probability models only tell us what we don't know. We never need, nor never should, use them for telling us what we already know. We know past observations and thus do not need probability models to report on observations. (Measurement error easily fits into this scheme: the real observations become what we don't know.)  Probability and statistical models cannot tell us about cause, and all those procedures which we currently use to imply cause should be abandoned.  Only the understanding of essence, nature, and powers tell us about cause.  Knowledge of causes is hard work, and the shortcuts provided by statistics are doing great harm. 

Even assuming we know the cause or partial causes of a set of observations, the standard methods of reporting, such as relative risk (odds ratios are similar), produce over-certainty, especially when reporting only on the parameters of probability models.  So-called predictive approaches compensate for this over-certainty, but they cannot eliminate it. 

\bibliographystyle{apalike}
\bibliography{/home/briggs/projects/writing/logic}

\end{document}